\documentclass[useAMS,usenatbib]{mn2e}

\usepackage{epsf}
\usepackage{amssymb}

\newcommand{\boojum}{\textsc{boojum}}

\newcommand{\kappaT}{\ensuremath{\kappa_{T}}}

\newcommand{\disc}{\ensuremath{\mathcal{D}}}
\newcommand{\contour}{\ensuremath{\mathcal{C}}}

\newcommand{\growth}{\ensuremath{\eta}}

\newcommand{\ncowl}{\ensuremath{\tilde{n}}}

\newcommand{\tdyn}{\ensuremath{\tau_{\rm dyn}}}

\newcommand{\Epul}{\ensuremath{E_{W}}}

\newcommand{\Lstar}{\ensuremath{L_{\ast}}}
\newcommand{\Rstar}{\ensuremath{R_{\ast}}}
\newcommand{\Mstar}{\ensuremath{M_{\ast}}}

\newcommand{\Lsun}{\ensuremath{{\rm L}_{\sun}}}
\newcommand{\Rsun}{\ensuremath{{\rm R}_{\sun}}}
\newcommand{\Msun}{\ensuremath{{\rm M}_{\sun}}}

\newcommand{\Xsurf}{\ensuremath{X_{\rm s}}}

\newcommand{\hr}{\ensuremath{\rm hr}}

\title[Excitation of g modes in Wolf-Rayet stars]
      {Excitation of g modes in Wolf-Rayet stars by a deep opacity bump}
\author[R. H. D. Townsend and J. MacDonald]
       {R. H. D. Townsend\thanks{Also at Bartol Research Institute}\thanks{E-mail: rhdt@bartol.udel.edu} and
        J. MacDonald\\
        Department of Physics \& Astronomy,
        University of Delaware,
        Newark, DE 19716, USA}

\date{%
Received: .................................... 
Accepted: ....................................
}

\pagerange{\pageref{firstpage}--\pageref{lastpage}}
\pubyear{2005}

\begin{document}


\maketitle

\label{firstpage}

\begin{abstract}
We examine the stability of $\ell=1$ and $\ell=2$ g modes in a pair of
nitrogen-rich Wolf-Rayet stellar models characterized by differing
hydrogen abundances. We find that modes with intermediate radial
orders are destabilized by a $\kappa$ mechanism operating on an
opacity bump at an envelope temperature $\log T \approx 6.25$. This
`deep opacity bump' is due primarily to L-shell bound-free transitions
of iron. Periods of the unstable modes span $\sim 11-21\,\hr$ in the
model containing some hydrogen, and $\sim 3-12\,\hr$ in the
hydrogen-depleted model. Based on the latter finding, we suggest that
self-excited g modes may be the source of the 9.8\,\hr-periodic
variation of WR~123 recently reported by \citet{Lef2005}.
\end{abstract}



\section{Introduction} \label{sec:intro}

Of all of the differing excitation mechanisms responsible for free
oscillations of stars \citep[for a comprehensive review, see][and
references therein]{Unn1989}, none is more ubiquitous than the
opacity, or `$\kappa$', mechanism. Put in simple terms, this mechanism
works in layers of a star's envelope where the opacity $\kappa$
increases in response to compression of the stellar material. Such an
increase (initially produced for instance by stochastic perturbations
to the equilibrium state) retards the escape of energy from the
stellar core, with the trapped flux being deposited in the layer as
heat. As \citet{Edd1926} discusses, the supply of additional heat
during compression creates a Carnot-cycle heat engine capable of
converting part of the outflowing radiant energy into mechanical
energy. In tandem with a restoring force that returns the layer back
toward its equilibrium state (pressure for p modes, buoyancy for g
modes), the heat engine creates a condition of pulsational instability
(overstability).

The role played by the $\kappa$ mechanism in the classical ($\delta$)
Cephei pulsators was first identified by \citet{Zhe1953}. In these
stars, the source of the instability is a peak in the opacity and its
partial derivative $\kappaT \equiv (\partial \ln \kappa/\partial \ln
T)_{\rho}$ at an envelope temperature $\log T \approx 4.65$ where
helium undergoes its second ionization. At lower luminosities, the
same mechanism is primarily responsible for the RR Lyrae and $\delta$
Scuti classes of pulsating star \citep[see][and references
therein]{Cox1980}.  However, in the early-type $\beta$ Cephei stars
and slowly pulsating (SPB) stars, the helium opacity bump is situated
too close to the stellar surface to play any significant role in
destabilizing pulsation modes.

In fact, the excitation mechanism responsible for these two classes of
star remained a mystery until the advent of new opacities from the
OPAL and OP projects \citep{RogIgl1992,Sea1994}. The treatment of
spin-orbit splitting in same M-shell transitions of iron and nickel,
absent in previous calculations, introduced a new opacity peak at
$\log T \approx 5.3$. Stability analyses by \citet{Cox1992},
\citet{DziPam1993}, and \cite{Dzi1993} revealed that this so-called
`iron bump' can destabilize p modes in $\beta$ Cephei stars and g
modes in SPB stars. More-recent investigations by \citet{Cha1997} and
\citet{Fon2003} have inferred that the same iron bump can also drive
the pulsation of the short-period (EC14026) and long-period subdwarf B
(sdB) stars.

In parallel with these developments, the new opacity data also
impacted understanding of the pre-white dwarf GW Vir (pulsating
PG1159) stars. \citet{Sta1983,Sta1984} found that g modes in these
objects could be $\kappa$-mechanism excited by an opacity bump around
$\log T \approx 6.2$, arising from K-shell photoionization of carbon
and oxygen; however, an almost-pure C/O mixture was required for the
instability to operate \citep{Sta1991}. This abundance restriction is
relaxed when updated OPAL/OP date are adopted \citep[see][and
references therein]{Gau2005}, due to the appearance of extra opacity
around $\log T \approx 6.3$. The origin of this additional opacity
appears to be spin-orbit effects in L-shell bound-free transitions of
iron \citep{RogIgl1992}.

From this brief survey of the literature, a natural question
emerges. The iron bump in the OPAL/OP opacities leads to instability
strips at both high masses ($\beta$ Cephei and SPB stars) and low
masses (sdB stars). Given that opacity bump at $\log T \sim 6-6.3$
drives pulsation in the low-mass GW Vir stars, is there by analogy a
high-mass group of stars that also exhibit pulsation driven by this
`deep opacity bump' (DOB)? In order for the pulsation to be remain
non-adiabatic around the driving zone, and therefore allow the
$\kappa$-mechanism heat engine to operate, these putative stars must
be exceedingly hot, with surface temperatures approaching
$10^{5}\,{\rm K}$. In fact, these stars can be recognized as massive,
population I, core helium burning Wolf-Rayet stars.

In this letter, we conduct a preliminary stability analysis of a pair
of model Wolf-Rayet (W-R) stars, to evaluate the hypothesis that g
modes can be excited in these stars by the DOB. We introduce the
models in Sec.~\ref{sec:models}, and describe the stability analysis
in Sec.~\ref{sec:stability}. Our results are presented in
Sec.~\ref{sec:results}, and then discussed and summarized in
Sec.~\ref{sec:discuss}.

\section{Stellar Models} \label{sec:models}

We evolve a star with an initial mass $M=100\,{\rm M}_{\sun}$ and
metalicity $Z=0.02$ using a stellar structure code written by JM. The
code is described in detail in \citet{Jim2004}; the details pertinent
to the present work are that (i) opacities are interpolated in the
revised OPAL tables by \citet{IglRog1996}, (ii) the heavy-element
mixture by \citet{GreNoe1993} is assumed for the initial composition,
and (iii) mass loss due to radiative driving is accounted for via a
modified form of the formula given by \citet{Abb1982}. The type-2 OPAL
tables used by the code allow variation of the abundances of carbon
and oxygen (in addition to hydrogen and helium), to reflect the
effects of nucleosynthesis. Opacity changes due to a varying nitrogen
abundance are simulated by adjusting the carbon and oxygen values
passed to the opacity interpolation routine. Since the CNO elements
are minor (and approximately interchangeable) opacity sources, this
procedure should not introduce any significant error.

We track the star's evolution off the main sequence, across to the red
region of the Hertzsprung-Russell (HR) diagram, and back again to the
blue. Two particular evolutionary stages are then selected as
exemplars of nitrogen-enriched (WN) Wolf-Rayet stars. The
``WNL''(late-type) model is somewhat hydrogen-depleted, with a surface
mass fraction $\Xsurf = 0.117$. The more-evolved ``WNE'' (early-type)
model is completely devoid of hydrogen, but has yet to show any
photospheric signature of carbon enrichment. Here, we use quotation
marks around our models' labels to emphasize that, in the same fashion
as \citet{Lan1994}, the designations `early' and `late' are to be
understood in an evolutionary rather than a spectroscopic sense.

\begin{table}
\leavevmode
\begin{centering}
\caption{Fundamental parameters of the stellar models introduced in
Sec.~\ref{sec:models}. Note that the radii \Rstar\ are for the
hydrostatic core, and do not include the extended wind region
containing the $\tau = 2/3$ photosphere \citep[see][]{Bas1991}.}
\label{tab:models}
\begin{tabular}{@{}cccccc}
model & age/Myr & $\log \Lstar/\Lsun$ & $\Mstar/\Msun$ &
$\Rstar/\Rsun$ & \Xsurf \\ \hline
``WNL'' & 3.57 & 5.77 & 20.6 & 10.3 & 0.117\\
``WNE'' & 3.74 & 5.62 & 16.2 & 2.10 & 0.000
\end{tabular}
\end{centering}
\end{table}

\begin{figure}
\leavevmode
\begin{centering}
\epsffile{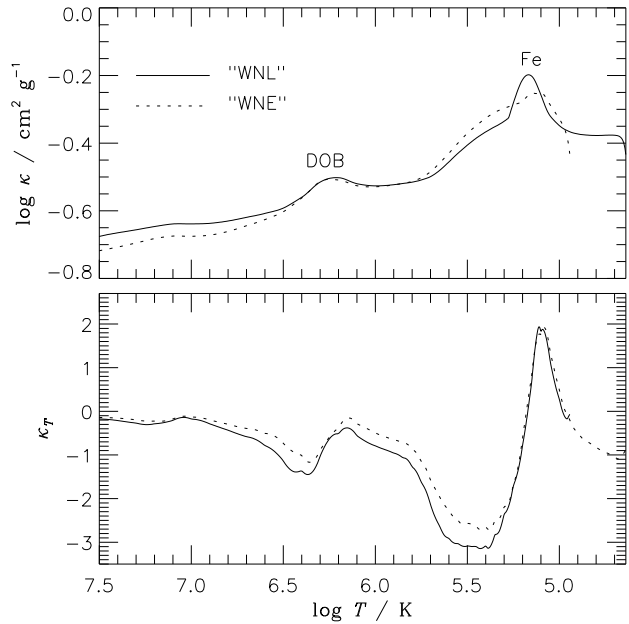}
\caption{The opacity $\kappa$ (top) and its temperature derivative
$\kappaT \equiv (\partial \ln \kappa/\partial \ln T)_{\rho}$ (bottom),
plotted as a function of envelope temperature for the ``WNE'' (solid)
and ``WNL'' (dotted) stellar models. The labels indicate the location
of the deep opacity bump (DOB) and iron opacity bump (Fe).}
\label{fig:kappa_T}
\end{centering}
\end{figure}

The fundamental parameters of both stellar models are summarized in
Table~\ref{tab:models}. In Fig.~\ref{fig:kappa_T}, we plot the opacity
and its derivative \kappaT\ as a function of temperature throughout
the model envelopes. The DOB can be seen clearly as the peak around
$\log T \approx 6.25$, while the iron bump also appears as a
pronounced peak in the superficial layers around $\log T \approx
5.15$. By exploring the effects of modifying the heavy-element mixture, we
have determined that the DOB is due primarily to iron, with minor
contributions coming from nickel and the CNO elements.

\section{Stability Analysis} \label{sec:stability}

To analyze the stability of the W-R stellar models introduced above,
we employ the \boojum\ non-radial, non-adiabatic pulsation code
developed by RHDT. The current revision incorporates a few improvements
over the baseline version described in \citet{Tow2005a,Tow2005b}.
Most significantly, the roots of the characteristic equation
\begin{equation}
\disc(\omega) = 0,
\end{equation}
that define the stellar eigenfrequencies, are now first isolated using
an approach similar to those suggested by \citet{Dzi1977} and
\citet{ShiOsa1981}. In this equation, $\omega$ is the pulsation
frequency normalized by the inverse dynamical timescale $\tdyn^{-1}
\equiv (G\Mstar/\Rstar^{3})^{1/2}$, and $\disc(\omega)$ is the
discriminant function defined by eqn.~(18) of \citet{Tow2005a}.

\begin{figure*}
\leavevmode
\begin{centering}
\epsffile{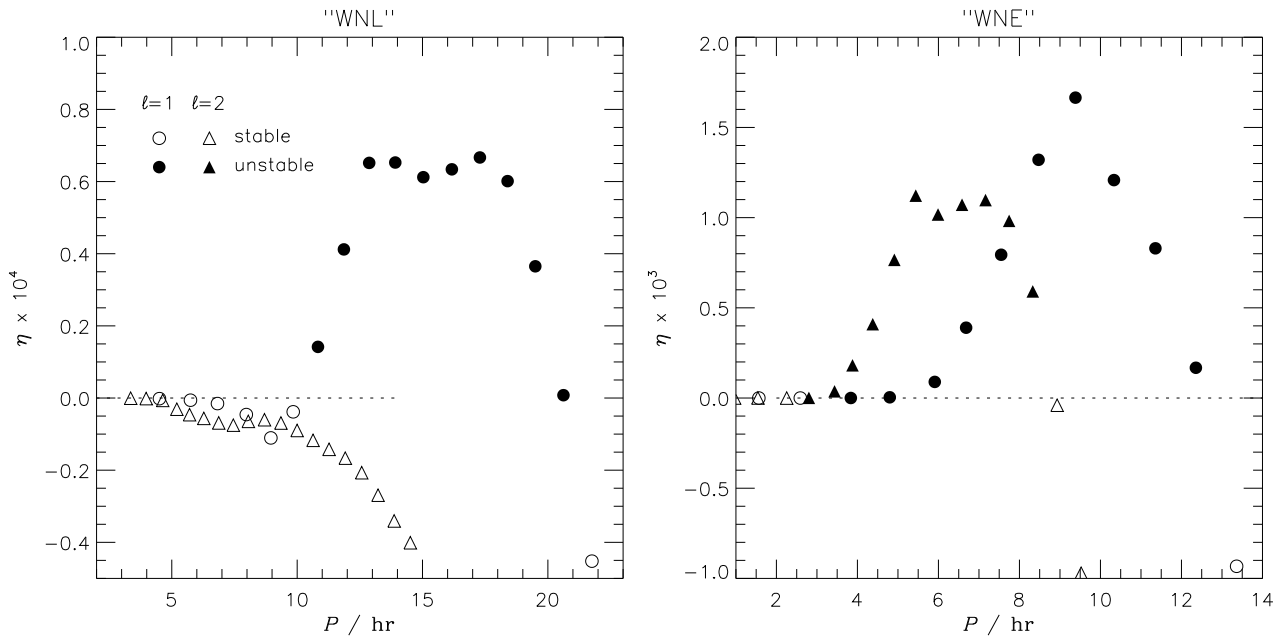}
\caption{Normalized growth rates \growth\ for the $\ell=1$ (circles) and
$\ell=2$ (triangles) modes of the ``WNL'' (left) and ``WNL'' (right) stellar
models, plotted as a function of pulsation period. The dotted line
indicates the division between stable modes ($\growth < 0$; open symbols) and
unstable modes ($\growth > 0$; filled symbols).}
\label{fig:growth}
\end{centering}
\end{figure*}

The root isolation proceeds by dividing the complex-$\omega$ plane
into many small rectangles, and evaluating for each one the contour
integral
\begin{equation}
\contour = \frac{1}{2\pi {\rm i}} \oint \frac{{\rm d}\disc(\omega)}{\disc(\omega)}
\end{equation}
counterclockwise around the perimeter, via a recursively refined
trapezoidal quadrature scheme. By Cauchy's theorem, \contour\
converges to the number of roots of the characteristic equation
enclosed by the rectangle. Once a root is thus isolated, the frequency
\begin{equation}
\omega_{\contour} = \frac{1}{2\pi {\rm i}
\,\contour} \oint \frac{\omega\,{\rm d}\disc(\omega)}{\disc(\omega)}
\end{equation}
is used as the starting point in the secant root-finding algorithm
implemented in \boojum. The \contour\ term in the denominator of this
expression ensures that $\omega_{0}$ lies within the enclosing
rectangle. Even though this term has an \emph{analytical} value of
unity, for a single enclosed root, its \emph{numerical} value differs
from unity due to the finite accuracy of our quadrature scheme.

This root isolation procedure is extremely useful in stars having high
luminosity-to-mass ratios (such as W-R and GW Vir stars), because the
pulsation tends to be so non-adiabatic \citep[see][their \S
22.1]{Unn1989} that solutions to the adiabatic pulsation equations
furnish very poor starting points for the secant root finder. However,
the contour integration often requires large numbers of quadrature
points before \contour\ converges satisfactorily toward an integer
value. Hence, although adding robustness to the process of solving the
non-adiabatic pulsation equations, root isolation is practical only
when -- as in the present work -- a few individual stellar models are
under study.

For the ``WNE'' and ``WNL'' models, we focus the stability analysis on
g modes having harmonic degrees $\ell=1$ and $\ell=2$, searching for
modes with radial orders \ncowl\ between, approximately, 0 and
-25. Here, \ncowl\ is defined by \citet[][their eqn.~17.5]{Unn1989},
within the generalization to the \citet{Cow1941} nomenclature
introduced by \citet{Scu1974} and \citet{Osa1975}. Negative values of
\ncowl\ typically correspond to g modes, but -- as is often the case
with more-evolved stellar models -- \ncowl\ does not vary
monotonically from one mode to the next, nor is it a unique index.

\section{Results} \label{sec:results}

The results from the stability analysis are presented in
Fig.~\ref{fig:growth}, which plots the normalized growth rate $\growth
\equiv -\Im(\omega)/\Re(\omega)$ of the modes found against their
period $P$. Evidently, the ``WNE'' model is unstable ($\eta > 0$)
against $\ell=1$ and $\ell=2$ g modes having periods in the range
$\sim 4-12\,\hr$ and $\sim 3-8\,\hr$, respectively. Similar
instability is seen in the ``WNL'' model for $\ell=1$ modes spanning
the period range $\sim 11-21\,\hr$; however, the $\ell=2$ modes of
this model are all found to be stable.

\begin{figure}
\leavevmode
\begin{centering}
\epsffile{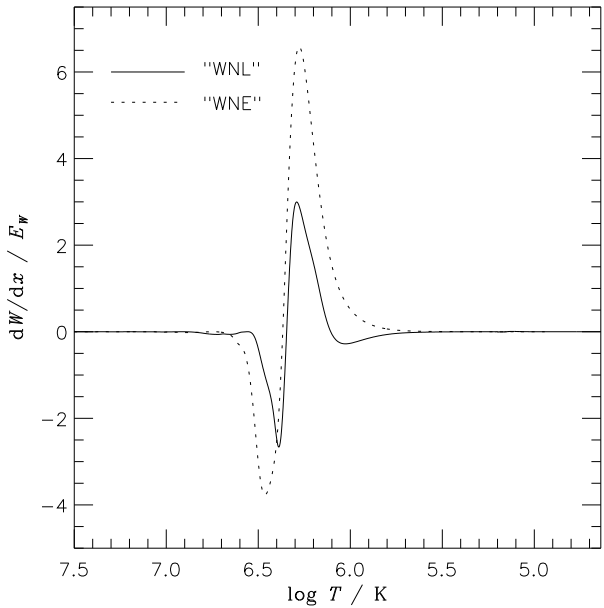}
\caption{Differential work functions, plotted as a function of
envelope temperature, for the $\ell=1$ modes of the two stellar models
having the largest growth rates. The differential is with respect to
the dimensionless radius $x\equiv r/\Rstar$, and the work $W$ is
expressed in units of the total energy of pulsation \Epul.}
\label{fig:work}
\end{centering}
\end{figure}

To explore the origin of the instability, Fig.~\ref{fig:work} plots
the differential work for the $\ell=1$ modes of each model that
exhibit the largest growth rates. As discussed by \citet[their \S
26.2]{Unn1989}, the $\kappa$ mechanism is operative wherever ${\rm
d}\kappa_{T}/{\rm d}r > 0$. From Fig.~\ref{fig:kappa_T}, we see that
this condition is met in the vicinity of the deep opacity bump, at a
temperature $\log T \approx 6.3$. Since this location coincides with
the sharp positive peak in both models' differential work (indicating
strong driving), we conclude that the unstable $\ell=1$ and $\ell=2$
modes are $\kappa$-mechanism excited by the DOB.

In Table~\ref{tab:growth}, we summarize the properties of the unstable
modes. The long-period limit of the instability arises because the
pulsation interior to the opacity bump, at a temperature $\log T
\approx 6.5$, becomes sufficiently non-adiabatic to activate strong
radiative damping there. Likewise, the short-period limit comes about
because low-order modes exhibit a relative Lagrangian pressure
perturbation $\delta p/p$ that is too small in the vicinity of the DOB
to generate appreciable excitation \citep[see][for a discussion of the
role played by $\delta p/p$]{Dzi1993}. The $\ell=2$ modes of the
``WNL'' model are all stable for much the same reason; their $\delta
p/p$ is peaked well exterior to the DOB, in a zone of inverted density
stratification situated in the outer envelope at $\log T \approx 5.2$.

This inversion zone, which is present in both models, is responsible
for a secondary family of unstable pulsation modes uncovered by
\boojum. These modes, not shown in Fig.~\ref{fig:work} due to our
choice of ordinate range, are found for harmonic degrees $\ell=1$ and
$\ell=2$, and are characterized by growth rates well in excess of
$10^{-2}$ and energy densities strongly concentrated in the inversion
zone. Following \citet{Sai1998}, we use this trapping property to
classify the modes as non-radial strange modes. We note that the iron
opacity bump is responsible for both the density inversions and the
instability of the strange modes trapped within them. Since the
strange modes are not related to the DOB, we shall not discuss them
further.

\section{Discussion} \label{sec:discuss}

In the preceding section, we demonstrate that low-degree, intermediate
radial-order g modes are unstable in a pair of stellar models
representative of WN stars, due to a $\kappa$ mechanism operating on
the deep opacity bump. This finding supports the central hypothesis of
the paper advanced in Sec.~\ref{sec:intro}, and likewise confirms WN
stars as the high-mass cousins of GW Vir stars, exhibiting a
relationship parallel to that between the high-mass $\beta$ Cephei and
SPB stars, and the low-mass sdB pulsators.

However, there are caveats to our results. Beyond the various
(reasonable) simplifications we adopt in constructing our stellar
models (Sec.~\ref{sec:models}) and conducting the stability analysis
(Sec.~\ref{sec:stability}), we have neglected the effects of the
strong, radiatively driven winds that are characteristic of all W-R
stars. The stellar models incorporate only the evolutionary effects of
wind-originated mass loss, and the stability analysis ignores the
radial outflow at the outer boundary. As discussed by \citet{Cra1996},
an outflow can allow ordinarily-trapped pulsation modes to propagate
outward through the boundary. \citet{Tow2000} demonstrated that this
leakage of wave energy can result in an appreciable damping of
pulsations. If the unstable g modes found herein are to remain
unstable, the rate of energy input from the $\kappa$ mechanism must
exceed the damping due to leakage.

A thorough evaluation of this issue may have to await further
theoretical advances in understanding how outflows influence
pulsation. In the meantime, we discuss an observational development of
particular relevance to the present work, one that poses some
interesting questions regarding the role played by pulsation (whether
leaking or not) in establishing structure and modulation in W-R
winds. Although searches for \emph{periodic} light variations in WN8
stars -- the subtype most characterized by variability -- have proven
largely unsuccessful \citep[see, e.g.,][and references
therein]{Mar1998}, recent \emph{MOST} observations of the WN8 star
HD~177230 (WR~123) by \citet{Lef2005} have detected an unambiguous
9.8\,\hr-periodic signal in the star's light curve.

Pulsation appears a promising candidate for the origin of the star's
variation. W-R stars have been known for some time to be unstable
toward strange modes \citep[e.g.,][]{Gla1993}; however, there is
debate over whether such modes are compatible with the observations of
WR~123. On the one hand, \citet{Lef2005} themselves argue that the
periods typical to strange modes, being on the order of the dynamical
timescale \tdyn, are far too short to match the 9.8\,\hr\ period
detected by \emph{MOST}. On the other hand, \citet{Dor2006} reason
that as a WN8 star WR~123 is expected to have a significantly larger
radius, and hence longer strange-mode periods, than the helium-star
models on which \citet{Lef2005} based their conclusions. In support of
their argument, \citet{Dor2006} present a WN8 model having
$\Rstar=15.4\Rsun$ that exhibits unstable radial strange modes with
periods on the order of a half day, seemingly compatible with the
observations.

\begin{table}
\leavevmode
\begin{centering}
\caption{Properties of the unstable $\ell=1$ and $\ell=2$ g modes of
the ``WNL'' and ``WNE'' stellar models. The subscripts $_{\rm min}$
and $_{\rm max}$ indicate the minimum and maximum values of the
relevant quantities, respectively.}
\label{tab:growth}
\begin{tabular}{@{}ccccccc}
model & $\ell$ & $P_{\rm min}$/\hr & $P_{\rm max}/\hr$ & $\ncowl_{\rm min}$ & $\ncowl_{\rm max}$ &
$\eta_{\rm max}$ \\ \hline
``WNL'' & 1 & 10.8 & 20.6 & -20 & -10 & $6.67 \times 10^{-5}$ \\
``WNL'' & 2 & \multicolumn{5}{c}{\emph{No unstable modes}} \\
``WNE'' & 1 & 3.83 & 12.4 & -14 & -4 & $1.67 \times 10^{-3}$ \\
``WNE'' & 2 & 2.81 & 8.34 & -15 & -5 & $1.12 \times 10^{-3}$
\end{tabular}
\end{centering}
\end{table}

Here, there may be a problem. The large radii of the models studied by
\citet{Dor2006} are, we believe, linked to the authors' \emph{ab
initio} assumption of an envelope hydrogen abundance $X =
0.35$. Although representative of many WN8 stars, this value is
inappropriate to WR~123, which is well-known to be largely depleted of
hydrogen \citep[$X \lesssim 0.005$; see, e.g.,][]{Cro1995}. It seems
probable that the absence of hydrogen in the star is a consequence of
it having already shed any residual hydrogen-rich envelope via wind
mass loss. Accompanying the loss of this envelope should be a
substantial reduction in the star's radius, as can be seen by
comparing our ``WNL'' and ``WNE'' models in Table~\ref{tab:models}.

Based on this reasoning, hydrogen-poor WR~123 may be characterized by
an appreciably smaller radius than assumed by \citet{Dor2006},
resulting once again in the period mismatch noted originally by
\citet{Lef2005}. This leads us to advance g modes excited by the DOB
as an alternative explanation for the star's variability. As
Fig.~\ref{fig:growth} illustrates, the 9.8\,\hr\ period is covered by
the unstable $\ell=1$ g modes of the hydrogen-poor ``WNE''
model. While we describe this model as `early-type', we once again
emphasize that the label is to be understood in an evolutionary rather
than spectroscopic sense.

The detectable presence of g modes in WR~123 could offer unparalleled
insights into the star's internal structure; these modes penetrate
down to the boundary of the convective core, and are therefore
particularly well suited to asteroseismological analyses. However, it
is unclear how best to proceed with the necessary initial step of
testing our interpretation of the observations. Pulsation in g modes
is often diagnosed through time-series monitoring of spectroscopic
line profile variations (lpv); the predominantly horizontal velocity
fields generated by these modes \citep[e.g.,][]{Unn1989} tend to
concentrate variability in the wings of line profiles. Unfortunately,
this approach is not feasible in the case of W-R stars, because the
supersonic wind washes out any lpv arising from (subsonic) pulsational
velocity fields. It remains to be seen whether other approaches can be
devised that are able to discriminate g modes from other potential
sources of variability.

Beyond the specific case of WR~123, the discovery of a new class of
Wolf-Rayet instability is of great interest in itself. Future
investigation can now focus on exploring the characteristics and
extent of the DOB instability in greater detail. Specific topics
worthy of attention include establishing red and blue edges in the HR
diagram, and determining the sensitivity of the instability against
changes in input physics (e.g., elemental diffusion, mass-loss rates,
etc.). As we have already noted, a proper treatment of pulsation at
the outflowing outer boundary is also needed. An integral part of this
treatment must be a better understanding of the reverse side of the
pulsation-wind interaction: how can pulsation at the wind base lead to
the modulations observed in Wolf-Rayet stars? Initial steps toward
addressing this question were made by \citet{CraOwo1996}, but there
remains much work to be done.


\section*{Acknowledgments}

We thank Stan Owocki and Tony Moffat for their helpful remarks. RHDT
acknowledges support from US NSF grant AST-0507581 and NASA grant
NNG05GC36G. He is also grateful for the Bickle.


\bibliography{deep}

\bibliographystyle{mn2e}


\label{lastpage}

\end{document}